\documentclass[aps,prl,preprint,superscriptaddress,showpacs]{revtex4}

\usepackage{amssymb}

\usepackage{epsfig}

\usepackage{amsmath}

\usepackage{times}

\begin{document}

\title{Geometric Structure of Two Self-dual Fields with Constraints}

\author{Xurong Chen}
\email{chen@physics.sc.edu}

\affiliation{Physics Department, University of South Carolina, Columbia, SC29208, USA}

\begin{abstract}
A two dimensional Poincar$\acute{e}$-invariant self-dual field with constraints is studied
in geometric way. We obtained its symplectic structure and conservative currents on  
space of solutions, which are also invariant under transformations of Poincar$\acute{e}$ group.

\end{abstract}

\date{\today}



\maketitle

\section{Introduction}   
Symplectic manifold associated with a 
time-dependent Lagrangian in classical mechanics is the space of motions. 
The correspondence in field theory is the infinite-dimensional symplectic manifold
of classical solutions of the field equations. Unfortunately, it gives a particular
canonical coordinate system which lets violate covariance because it 
selects out a particular coordinate system. It is not inherent in the canonical approach. 
In geometric approach, the coordinates are irrelevant since all we just need 
symplectic structure~\cite{woodhouse}. The symplectic structure can be obtained directly from
the Lagrangian without violating covariance. 

The geometric approach is a powerful 
tool to study classical mechanics systems and their symmetry. For a given 
classical field, space of motions is its 
phase space~\cite{witten}. Symplectic structures on space of motions
can be obtained from its Lagrangian. From Poincar$\acute{e}$ invariant
symplectic structure is, the classical field's conservative
currents can be calculated. Hence we can give out geometric descriptions
of classical fields.

In this paper, we will apply the above approach to self-dual field.  
self-dual fields $\phi(x,t)$, sometimes called chiral bosons, satisfy the
self-dual condition,
\begin{equation} \label{eq:selfdual}
\dot{\phi} = \phi^{\prime},
\end{equation}
where the overdot denotes differentiation with respect to time t and prime to
space x. The self-dual fields have been studied widely.
Yang and Luo~\cite{yang} discussed geometric structure for free two-dimensional 
self-dual field. In this paper self-dual field with constraints is studied. 
By symplectic technique, we obtained its geometric description.   

\section{Hamiltonian description}  
In this paper we study the following Lagrangian density with 
constraints ~\cite{kulshreshtha}
\begin{equation} \label{eq:selfduallag}
\mathcal{L} = \frac{1}{2}\dot{\phi}^2 - \frac{1}{2}\phi^{\prime 2} +
\lambda(\dot{\phi} - \phi^{\prime}),
\end{equation}
where $\lambda = \lambda(x,t)$ is an auxilary field.
The equations of motions are
\begin{equation} \label{eq:selfdualeq1}
\dot{\phi} = \phi^{\prime},\qquad
\dot{\lambda} = \lambda^{\prime}.
\end{equation}
So it is a self-dual field.

Firstly let us study the canonical treatment of this system. 
To otbain the canonical formalism, we should treat $\phi$,
$\lambda$ and their corresponding conjugate canonical momenta
$\pi$, p at each point (t,x)
as independent coordinates on the momentum phase space V~\cite{kulshreshtha}.
\begin{equation}
\pi = \frac{\partial\mathcal{L}}{\partial(\partial_t\phi)} = \dot{\phi} +\lambda,
\end{equation}
\begin{equation}\label{eq:p1}
p = \frac{\partial\mathcal{L}}{\partial\dot{\lambda}} = 0,
\end{equation}
Eq.(~\ref{eq:p1}) is a primary constraint,
\begin{equation}
\Omega_1: = p \approx 0.
\end{equation}
The secondary constraint is 
\begin{equation}
\Omega_2: = \partial_t \Omega_1 = \pi - \lambda - \phi^{\prime} \approx 0.
\end{equation}
With these two constraints, reduction of phase space becomes to have only 
two degrees of freedom. If we can choose $(\pi, \phi)$ as two independent
variables, Hamiltonian is
\begin{equation}
\mathcal{H} = \pi\dot{\phi} - \mathcal{L} = \frac{1}{2}(\pi - \lambda)^2 +
\frac{1}{2}\phi^{\prime 2} + \lambda\phi^{\prime} = \pi \phi^{\prime}.
\end{equation}
The Poisson brackets among the constraints are
\begin{equation}
\{\Omega_1(x,t), \Omega_1(y,t)\} = 0,
\end{equation}
\begin{equation}
\{\Omega_1(x,t), \Omega_2(y,t)\} = \delta(x - y),
\end{equation}
\begin{equation}
\{\Omega_2(x,t), \Omega_2(y,t)\} = -2\partial_1\delta(x - y).
\end{equation}
From these commutation relations, Dirac brackets $\{, \}_D$ were calculated,
and then quantization were discussed in ~\cite{kulshreshtha}.

\section{Symplectic structure}  
For any two vectors $\psi = (\psi^{\alpha})$ and $\bar{\psi} =
(\bar{\psi^{\beta}})$ on space of solutions $\mathcal{M}$, symplectic structure on $\mathcal{M}$ is~\cite{woodhouse}
\begin{equation} 
\omega(\psi,\bar{\psi}) = \frac{1}{2}\int_{\Sigma}\omega^{\nu}d\sigma_{\nu},
\end{equation}
where $d\sigma_{\nu} = n_{\nu}d\sigma$; $d\sigma$ is the volume element on
$\Sigma$ and $n_{\nu}$ is the unit normal. $\Sigma$ $\subset$ Q is a spacelike
hypersurface in Minkowski space. $\omega = d\theta$ is independent
of $\Sigma$ and is a natural closed 2-form on $\mathcal{M}$~\cite{woodhouse}.
\begin{equation}
\omega^{\nu} = \frac{\partial^2\mathcal{L}}{\partial\psi^{\beta}\partial\psi_{\nu}^{\alpha}}(\psi^{\beta}\bar{\psi}^{\alpha} - \bar{\psi}^{\beta}\psi^{\alpha}) + \frac{\partial^2\mathcal{L}}{\partial\psi_{b}^{\beta}\partial\psi_{\nu}^{\alpha}}(\bar{\psi}^{\alpha}\nabla_b\psi^{\beta} - \psi^{\alpha}\nabla_b\bar{\psi}^{\beta}), \qquad
\nu = 0, 1.
\end{equation},
In our case, 
$\psi = (\phi, \lambda),$
$\bar{\psi} = (\bar{\phi}, \bar{\lambda}).$
Its symplectic structure
\begin{equation}
\omega^{0} = (\lambda\bar{\phi} - \bar{\lambda}\phi) + (\bar{\phi}\dot{\phi} - \phi\dot{\bar{\phi}}),
\end{equation}
\begin{equation}
\omega^{1} = (\bar{\lambda}\phi - \lambda\bar{\phi}) + (\phi\bar{\phi}^{\prime} - \bar{\phi}\phi^{\prime}),
\end{equation}
we have
\begin{equation}
\partial_{\nu}\omega^{\nu} = \partial_t\omega^0 + \partial_x\omega^1 = 0. 
\end{equation}
So $\omega^{\nu}$ ($\nu$ = 0, 1) are conservative symplecitc currents.

Now let us consider two dimensional conformal transformation,
\begin{equation}
\delta_fx_{\mu} = x_{\mu}^{\prime} - x_{\mu} = f_{\mu},
\end{equation}
$\mu = 0, 1$, which is larger than the Poincar$\acute{e}$ group of translations and 
Lorentz rotations~\cite{jackiw}. Conformal Killing vector equation for this transformation is  
\begin{equation}
\partial_{\mu}f_{\nu} + \partial_{\nu}f_{\mu} = \partial_{\lambda}f^{\lambda}g_{\mu\nu}.
\end{equation}
Its Lie agebra
\begin{equation}
\lbrack\delta_f, \delta_g\rbrack = \delta_{(f,g)},
\end{equation}
where $(f, g)_{\mu} = f^{\alpha}\partial_{\alpha}g_{\mu} - g^{\alpha}\partial_{\alpha}f_{\mu}$,
$\alpha = 0, 1$. For self-dual field $\phi(x,t)$ we have 
\begin{equation}
\delta_f\phi = f^{\mu}\partial_{\mu}\phi.
\end{equation}
\begin{equation} 
\Delta\omega = \int_{\Sigma}\delta\omega^{\nu}d\sigma_{\nu},
\end{equation}
\begin{equation} 
\Delta\omega^0 = (f^0 + f^1)(\dot{\lambda}\bar{\phi} + \lambda\dot{\bar{\phi}} - \dot{\bar{\lambda}}\phi - \bar{\lambda}\dot{\phi} + \ddot{\phi}\bar{\phi}
                - \phi\ddot{\bar{\phi}}); 
\end{equation}
\begin{equation} 
\Delta\omega^1 =(f^0 + f^1)(\dot{\bar{\lambda}}\phi + \bar{\lambda}\dot{\phi} - \dot{\lambda}\bar{\phi} - \lambda\dot{\bar{{\phi}}} + \phi\dot{\bar{\phi}}^{\prime} - \dot{\phi}^{\prime}\bar{\phi}). 
\end{equation}
if we require
\begin{equation} 
\Delta\omega = 0,
\end{equation}
then there must be
\begin{equation} 
f^0 + f^1 = 0.
\end{equation}
It means the two-dimensional conformal group contracts to one component.
This result is consistent with  ~\cite{jackiw1986}.

\section{Conservative currents} 
Generaly, for a given Lagrangian system, according to Noether theorem,
we can study its concervative currents with symmetries of the Lagrangian
by geometric approach.

let Q be background space-time, there is a Killing vector V(x) on it generating
a flow $\rho_t$
\begin{equation}
\rho_t = Q \to Q.
\end{equation}
\begin{equation}
\frac{d}{dt}|_{t=0}\rho_t(x) = V(x), \qquad x \subset Q. 
\end{equation}
$\rho_t$ induces a ``push forward'' operator $R_t$
\begin{equation}
R_t(\phi) = \phi \cdot \rho_t^{-1},
\end{equation}
If $\phi$ is a solution of a field equation, then so is $R_t \phi$. 
So $R_t$ maps solutions to solutions, and hence induces a flow on $\mathcal{M}$\cite{woodhouse}.
We call $R_t$ is hamiltonian current on the manifold $\mathcal{M}$, which preserves
Hamilton. Hamiltonian vector field $X$ is defined by:
\begin{equation}
\frac{d}{dt}(R_t \phi^{\alpha})_{t=0}  = X^{\alpha},
\end{equation}
and conservative currents are
\begin{equation}
J_{\mu} = X(\phi)\frac{\partial \mathcal{L}}{\partial(\partial_{\mu}\phi)} +
V^{\mu}(x)\mathcal{L}.
\end{equation}

For the Poincar$\acute{e}$-invariant self-dual field, it is easy to obtain 
its various conservative geometric currents:
\begin{itemize}
\item We first consider time and space translations.
Conserved quantities are total energy and total momentum:
\begin{equation}
H = \int(\frac{1}{2}\dot{\phi}^2  \frac{1}{2}\phi^{\prime 2} + \lambda\phi^{\prime})dx.
\end{equation}
\begin{equation}
P = - H.
\end{equation}

\item For Lorentz transformations
\begin{equation}
x_{\mu}^{\prime} = a_{\mu\nu}x_{\nu}  
\qquad \mu, \nu = 0,1,
\qquad a_{\mu\nu} = \delta_{\mu\nu} + \alpha_{\mu\nu}, 
\end{equation}
$\alpha_{\mu\nu}$ is a infinite asymmetry. Its Killing vector 
\begin{equation}
V^{\mu}(x) = \frac{\partial}{\partial s}|_s\alpha_{\mu\nu}(s)x_{\nu},
\qquad \beta_{\mu\nu} \equiv \frac{\partial}{\partial s}|_s\alpha_{\mu\nu},
\end{equation}
and Hamiltonian vector
\begin{equation}
X(\phi) = \frac{d}{ds}|_{s=0}\phi^{\prime}(\rho_sx) =
\partial_{\lambda}\phi\beta_{\lambda\nu}x_{\nu}.
\end{equation}
It generates conservative currents
\begin{equation}
J_{\mu} =
\beta_{\lambda\nu}x_{\nu}\partial_{\lambda}\phi\frac{\partial\mathcal{L}}{\partial(\partial_{\mu}\phi)} + \beta_{\mu\nu}x_{\nu}\mathcal{L}
= \frac{1}{2}\beta_{\mu\nu}(x_{\nu}T_{\mu\lambda} - x_{\lambda}T_{\mu\nu}) =
\frac{1}{2}\beta_{\mu\nu}L_{[\lambda\nu]\mu},
\end{equation}
where $L_{[\lambda\nu]\mu}$ is the 3rd orbital angular momentum tensor. The $J^{\mu}$
corresponds to total orbital angular momentum.

\item Since the self-dual field is invariant under one-parameter 
two-dimensional conformal transformations (with condition $f^0 + f^1 = 0$),
so we also can study conservative currents it generates.
\begin{equation}
x_{\mu}^{\prime} = x_{\mu} + f_{\mu}(x(s)),
\end{equation}
\begin{equation}
V^{\mu}(x) = \partial_{\nu}f_{\mu}\frac{d}{ds}|_{s=0}x_{\nu}(s) +
\frac{d}{ds}|_{s=0}x_{\mu}(s) = (\delta_{\mu\nu} + \partial_{\nu}f_{\mu})u_{\nu},
\end{equation}
where $u_{\nu} \equiv \frac{d}{ds}|_{s=0}x_{\nu}(s)$.
\begin{equation}
X(\phi) = \frac{d}{ds}|_{s=0}\phi^{\prime}(\rho_sx) =
\partial_{\lambda}\phi u_{\lambda},
\end{equation}
\begin{equation}
J_{\mu} =
\partial_{\lambda}\phi u_{\lambda}\frac{\partial\mathcal{L}}{\partial(\partial_{\mu}\phi)} + (\delta_{\mu\nu} + \partial_{\nu}f_{\mu})u_{\nu}\mathcal{L}
= u_{\lambda}T_{\lambda\mu} + \partial_{\nu}f_{\mu}u_{\nu}\mathcal{L}.
\end{equation}
\end{itemize}

\section{Conclusions} 
We study Poincar$\acute{e}$-invariant two dimensional self-dual field with constraints. 
Symplectic structure and conservative geometric currents on 
space of motions were obtained. It made the first step for geometric quantization of self-dual
fields.

\section{Acknowledgments} 
We would like to thank K. Q. Yang and Y. Luo for stimulating discussions and many insightful comments.

\end{document}